# Training and Comparison of nnU-Net and DeepMedic Methods for Autosegmentation of Pediatric Brain Tumors

Arastoo Vossough[1,2,3], Nastaran Khalili[1], Ariana M. Familiar[1], Deep Gandhi [1], Karthik Viswanathan[1], Wenxin Tu[4], Debanjan Haldar[1], Sina Bagheri[1,2], Hannah Anderson[1], Shuvanjan Haldar [5], Phillip B. Storm[1,6], Adam Resnick[1], Jeffrey B. Ware[2], Ali Nabavizadeh[1,2*], Anahita Fathi Kazerooni[1,6,7*]

* equally-contributing last authors

[1] *Center for Data Driven Discovery in Biomedicine (D³b), Children's Hospital of Philadelphia, Philadelphia, PA, USA*

[2] *Department of Radiology, University of Pennsylvania, Philadelphia, PA, USA*

[3] *Department of Radiology, Children's Hospital of Philadelphia, Philadelphia, PA, USA*

[4] *College of Arts and Sciences, University of Pennsylvania, Philadelphia, PA, USA*

[5] *School of Engineering, Rutgers University, New Brunswick, NJ, USA*

[6] *Department of Neurosurgery, Children's Hospital of Philadelphia, Philadelphia, PA, USA*

[7] *Center for AI & Data Science for Integrated Diagnostics (AI²D) and Center for Biomedical Image Computing and Analytics (CBICA), University of Pennsylvania, Philadelphia, PA, USA*

**Corresponding author:**

Anahita Fathi Kazerooni, PhD

Center for Data Driven Discovery in Biomedicine (D³b), Children's Hospital of Philadelphia, Philadelphia, PA, USA

Email: anahitaf@upenn.edu; fathikazea@chop.edu

Funding:

This research was supported by grant funding from the DIPG/DMG Research Funding Alliance (DDRFA) and Pediatric Brain Tumor Foundation (PBTF).

**ABSTRACT**

Background and Purpose: Brain tumors are the most common solid tumors and the leading cause of cancer-related death among children. Tumor segmentation is essential in surgical and treatment planning, and response assessment and monitoring. However, manual segmentation is time-consuming and has high inter-operator variability, underscoring the need for more efficient methods. We compared two deep learning-based 3D segmentation models, DeepMedic and nnU-Net, after training with pediatric-specific multi-institutional brain tumor data using based on multi-parametric MRI scans.

Materials and Methods: Multi-parametric preoperative MRI scans (pre- and post-contrast T1WI, T2WI, and FLAIR) of 339 pediatric patients (n = 293 internal and n = 46 external cohorts) with a variety of tumor subtypes, were preprocessed and manually segmented into four tumor subregions, i.e., enhancing tumor (ET), non-enhancing tumor (NET), cystic components (CC), and peritumoral edema (ED). After training, performance of the two models on internal and external test sets was evaluated using Dice scores, sensitivity, and Hausdorff distance with reference to ground truth manual segmentations. Additionally, concordance was assessed by comparing the volume of the subregions as a percentage of the whole tumor between model predictions and ground truth segmentations using Pearson's or Spearman's correlation coefficient and the Bland-Altman method.

Results: Dice score for nnU-Net internal test sets was (mean ± SD (median)) 0.9±0.07 (0.94) for WT, 0.77±0.29 for ET, 0.66±0.32 for NET, 0.71±0.33 for CC, and 0.71±0.40 for ED, respectively. For DeepMedic the Dice scores were 0.82±0.16 for WT, 0.66±0.32 for ET, 0.48±0.27, for NET, 0.48±0.36 for CC, and 0.19±0.33 for edema, respectively. Dice scores were significantly higher for nnU-Net (p≤0.01). Correlation coefficients for tumor subregion percentage volumes were higher and Bland-Altman plots better for nnU-Net compared to DeepMedic. External validation of the trained nnU-Net model on the multi-institutional BraTS-

PEDs 2023 dataset revealed high generalization capability in segmentation of whole tumor and tumor core with Dice scores of 0.87 ± 0.13 (0.91) and 0.83 ± 0.18 (0.89), respectively.

Conclusion: Pediatric-specific data trained nnU-Net model is superior to DeepMedic for whole tumor and subregion segmentation of pediatric brain tumors.

**Abbreviations:**

AI = artificial intelligence; CBTN = Children's Brain Tumor Network; WT = whole tumor; ET = enhancing tumor; NET = nonenhancing tumor; CC = cystic component; ED = edema; TC = tumor core; FLAIR = fluid attenuated inversion recovery; CNN = Convolutional Neural Networks.

## INTRODUCTION

Pediatric CNS tumors are the second most common childhood cancer and represent the most prevalent solid tumor and the leading cause of cancer-related mortality in children [1, 2]. These tumors encompass a wide range of histologies and display significant variations in their molecular origins, disease course, and response to therapy, which complicates the clinical decision-making process for their management [3]. MRI scans are imperative to precisely locate, characterize, and monitor the treatment for these tumors. Achieving this precision heavily relies on accurate delineation and characterization of the whole tumor and tumor subcomponents. Quantitative measures of change in tumors are highly desirable for objective assessments of size and signal intensity. This requires visual and manual measurement of the tumors and tumor components. Accurate, automated tumor segmentation methods can offer rapid determination of tumor volumes with less effort and potentially more consistency than manual segmentation methods.

Manual delineation of brain tumors presents distinct obstacles that require specialized expertise, resources, and time [4]. Moreover, there are distinct differences in the prevalence, appearance, histology, and behaviors of pediatric brain tumors compared to adult brain tumors [1,5]. The most common pediatric brain tumors include pilocytic astrocytoma, medulloblastoma, and other gliomas. In adults, the most common intracranial tumors are brain metastases and meningiomas and the most common primary intra-axial brain tumors are glioblastoma. There is a higher prevalence of circumscribed gliomas in children compared to infiltrating gliomas that are commonly seen in adults. Necrosis is much more prevalent in adult brain tumors than in children, whereas tumoral cysts are more common in pediatrics. Contrast-enhancement in adult brain tumors is more commonly associated with high-grade tumors whereas a large proportion of low-grade pediatric tumors demonstrate contrast-enhancement. The response assessment criteria for adult and pediatric brain tumors are also significantly different [6-8]. The differences have implications for brain tumor subregion segmentation. In adult brain tumor segmentations, typically nonenhancing tumor and edema are combined into a single subregion label given the difficulty of separating these tissues and common infiltrative adult brain tumors. Alternatively, nonenhancing tumor and tumor necrosis may be combined into one label, although they are quite distinct on imaging. As a result of these differences, segmentation models trained on adult brain tumors may not be well-suited for segmentation of pediatric brain tumors, leading to under- or over-segmentation of tumor subregions [9]. As such, there is a relatively unmet need for training and validation of more accurate, pediatric-specific, brain tumor segmentation models with tumor subregion delineation.

Advancements in deep learning have significantly broadened the potential applications of artificial intelligence (AI) in the field of medical imaging. In the context of brain tumors, 3D Convolutional Neural Networks (CNNs) have gained widespread use in this field for image segmentation due to their ability to capture spatial features in 3D data, which is particularly relevant for brain MRI scans [10]. Despite this, optimal utilization of CNNs for pediatric brain tumor segmentation remains inadequate [11]. Distinctive characteristics of the pediatric brain as well as inherent limitations in available data sources present formidable challenges in the development of automated segmentation methods tailored to pediatric cases [12]. Moreover, given the necessity for AI systems to undergo specific training and validation for each distinct application, it is evident that focused research endeavors within the pediatric demographic are essential for optimal application of AI in segmenting pediatric brain tumors.

Recent advancements in CNN structures like ShuffleNet, ResNet, and DenseNet have shown promising results in various image analysis tasks, including adult brain tumor segmentation [13, 14]. U-Net, a CNN based on an encoder-decoder architecture, has become popular for medical image segmentation, displaying superior performance in small datasets with a limited number of scans [15, 16]. As such, U-Net and its derivatives are potentially helpful for segmentation of pediatric brain tumors. A few studies have used U-Net models to improve segmentation performance on particular types of pediatric brain tumors, yielding good agreement between predicted and manual segmentations [17, 18]. Nevertheless, there is a paucity of available highly accurate deep learning models that can be used for pediatric brain tumor segmentation across a wide range of tumor pathologies.

In this paper, we investigate two innovative 3D deep learning segmentation architectures that have been successfully applied in adult brain tumor segmentations, namely DeepMedic and nnU-Net models [19, 20]. Our objective was to harness an extensive collection of multi-institutional ground truth segmentations to train these two models on carefully curated, pediatric-specific data and compare the performance for automated pediatric brain tumor subregion segmentation across a wide spectrum of tumor types.

**MATERIALS AND METHODS**

**Data Description and Patient Cohort**

This was a HIPAA-compliant, IRB-approved study of previously acquired multi-institutional data from the subjects enrolled onto the Children's Brain Tumor Network (CBTN) consortium (https://cbtn.org) [21]. MRI exams of pediatric patients with histologically confirmed brain tumors from the CBTN consortium were retrospectively collected. Inclusion criteria comprised of availability of preoperative brain MRI comprising four conventional MRI sequences, i.e., pre-contrast T1WI, T2WI, T2-Fluid Attenuated Inversion Recovery (FLAIR) and gadolinium post-contrast T1WI (T1WI-Gd) sequences, all acquired as a part of standard-of-care clinical imaging evaluation for brain tumors. Patients were still included if the only procedure was placement of an external ventricular drain or needle biopsy. Patients were excluded if the images were incomplete or if severely degraded by artifacts. Internal site data was from the XXXX hospital and external data was from other consortium members of CBTN. A total of 339 patients (293 from internal site, 46 from external sites) were included in this study. Detailed descriptions of the patients, tumor types, and MRI scans characteristics are included in Table 1 and Supplementary Table 1.

Details about image preparation, preprocessing, and tumor subregion segmentation can be found in the Supplementary Material. Tumors were segmented into four subregions [9], including enhancing tumor (ET), non-enhancing tumor (NET), cystic component (CC), and peritumoral edema (ED). Whole tumor (WT) segmentation masks were generated by union of all four tumor components (i.e., WT = ET + NET + CC + ED).

**Model Training and Validation**

We trained and evaluated two 3D convolutional neural networks, DeepMedic and nnU-Net, for automated tumor subregion segmentation on multiparametric MRI sequences of 233 subjects from the internal cohort and tested on withheld sets of 60 internal and 46 external subjects. nnU-Net v1 (https://github.com/MIC-DKFZ/nnUNet/tree/nnunetv1) with 5-fold cross-validation was trained with an initial learning rate of 0.0, stochastic gradient descent (SGD) with Nesterov momentum ($\mu = 0.99$), and number of epochs = 1000 x 250 minibatches. As the DeepMedic approach does not inherently include cross-validation, a validation set comprising 20% of the 293 training subjects (n = 47) was randomly selected. DeepMedic v0.8.4 (https://github.com/deepmedic/deepmedic) was trained from scratch with a learning rate = 0.001, number of epochs = 35, and batch size = 10.

**Performance Analysis**

The performance of the DeepMedic and nnU-Net models with respect to the expert manual ground truth segmentations were evaluated using several evaluation metrics, including Dice score (Sørensen-Dice similarity coefficient), sensitivity, and 95% Hausdorff distance. We separately assessed segmentation of WT, ET, NET, CC, and ED subregions along with the non-enhancing component/edema, which encompassed the combination of NET, CC, and ED. Further evaluation of correlation between model predictions (automated) and ground truth (manual) segmentations was done by calculating the percentage volume of the subregions to the whole tumor between ground truth and automated models using Pearson's or Spearman's correlation coefficient, depending on the data distribution. Additionally, the agreement between segmentations predicted by the nnU-Net and DeepMedic models in relation to the ground truth segmentations was examined using the Bland-Altman method. We calculated the proportion of the whole tumor volume that is ET, NET, CC, or ED subregions for the two models and compared them to the ground truth for each of the two models using the Mann-Whitney U-test.

**Benchmarking Our Model in the BraTS-PEDs Context**

We extended the validation of our nnU-Net model to include the latest benchmarks in automated tumor segmentation, specifically focusing on the multi-institutional dataset provided through Brain Tumor Segmentation Challenge in Pediatrics (BraTS-PEDs 2023 dataset [22, 23]). Our analysis involved applying the nnU-Net model to a cohort of 92 pediatric subjects diagnosed with high-grade gliomas, which included astrocytoma and diffuse midline glioma/diffuse intrinsic pontine glioma (DMG/DIPG).

The evaluation of our model's performance was conducted in alignment with the BraTS-PEDs validation criteria [22]. We focused on the segmentation of three key areas: the enhancing tumor (ET), the tumor core (TC), and the whole tumor (WT). The TC encompassed the ET, NET, and CC regions.

**Code Availability**

All image processing tools used in this study are freely available for public use (CaPTk, https://www.cbica.upenn.edu/captk; ITK-SNAP, https://www.itksnap.org). The pre-trained nnUNet tumor segmentation model is publicly available on GitHub (https://github.com/d3b-center/peds-brain-auto-seg-public). It is also a software plug-in ("gear") on the Flywheel platform and can be found by searching "Pediatric Brain Automated Segmentation" in the Flywheel Gear Exchange library.

## RESULTS

The Dice score, sensitivity, and 95% Hausdorf distance metrics of the nnU-Net and DeepMedic deep learning models compared to manual ground truth segmentations are shown in Table 2, and Supplementary Tables 2-3. Median values are included in addition to mean and standard deviation since a discrepancy in the mere absence or presence of even a small tumor subregion label in one of the model-pair comparisons may tremendously affect the metrics for a particular subject and disproportionately affect the mean (e.g., result in a calculated Dice score of 0). Additionally, the distribution of the internal and external tests set Dice scores of whole tumor and tumor subregions for the two models is shown with violin plots in Figure 1. The distribution of Dice scores in whole tumor and across all tumor subregions was more favorable for nnU-Net compared to DeepMedic, with a tighter distribution towards higher Dice scores. Both models performed worse in ED segmentation compared to other tumor subregions, but even here nnU-Net performed much better than DeepMedic, which had rather poor results for edema delineation (Dice score = 0.19 and 0.21 for internal and external test subjects, respectively). Paired t-test comparison of Dice scores between the two automated segmentation methods demonstrated higher nnU-Net Dice scores compared to DeepMedic for enhancing tumor in the internal test set ($p=0.01$) and also for whole tumor and all other tumor subregions in both internal and external test sets ($p<0.001$).

Correlation between the volume percentages of tumor subregion segmentation compared to the ground truth for nnU-Net and DeepMedic models are shown in Figure 2, Supplementary Figure 1, and Supplementary Table 4. All correlation coefficient p-values were less than 0.001, but again the lowest correlations were seen with DeepMedic ED determination, with $r = 0.48$ and $r = 0.33$ for internal and external test subjects, respectively. All nnU-Net percentage volume correlation coefficients were close to or above $r = 0.9$. Furthermore, Bland-Altman assessment of agreement between the methods showed tighter 95% intervals for nnU-Net when compared to the ground truth as opposed to DeepMedic (Figure 3 and Supplementary Figure 2).

Finally, results of Mann-Whitney U test comparing tumor subregions as a proportion of whole tumor in nnU-Net and DeepMedic segmentations compared to ground truth segmentation are shown in Supplemantary Table 5. Edema (ED) segmentation proportions determined by DeepMedic were statistically different from the ground truth for the internal test subjects. For the internal test subjects, the DeepMedic model was significantly different from ground truth in the proportion of ED with respect to the whole tumor.

Sample MRI images with results comparing ground truth segmentation to nnU-Net and DeepMedic models are showcased in Figure 4 and Supplementary Figures 3-4.

Our analysis of the nnU-Net-based autosegmentation model applied to the multi-institutional BraTS-PEDs 2023 dataset (Table 3), indicates a high degree of generalizability in segmenting both the whole tumor and the tumor core regions. However, a comparatively lower performance in segmenting the enhancing tumor region was observed.

**DISCUSSION**

DeepMedic is a multi-layered, multi-scale 3D deep CNN architecture coupled with a 3D fully connected conditional random field [19], showing excellent performance in adult brain tumors, notably in the 2017 BraTS challenge by the Medical Image Computing and Computer Assisted Interventions (MICCAI) organization. It has been effective in pediatric tumor subregion segmentation [9], and has outperformed technicians in brain tumor segmentation on a multi-institutional MRI database, as validated by neuroradiologists [24]. Its multi-scale, parallel processing approach captures comprehensive contextual information. Utilizing feature maps from its final convolutional layer, DeepMedic efficiently predicts voxel labels in input patches [25]. However, its focus on high-level semantic information may limit its performance [25].

nnU-Net, a self-configuring model for biomedical image segmentation, excels in various applications [20], including the BraTS 2020 multiorgan segmentation challenge [26]. It has also been successfully applied in automated segmentation of brain metastases and specific tumors such as craniopharyngiomas [27] or meningiomas [28]. Based on the U-Net structure, it combines encoding with downsampling and decoding with upsampling. U-Net introduces multilevel information gradually from encoding to decoding to optimize prediction accuracy. nnU-Net extends the original U-Net design with an automated end-to-end pipeline, selecting the optimal configuration for diverse segmentation tasks.

Leveraging accurate automated methods can greatly facilitate the rapid delineation of structures and potentially decrease variability. To this end, we leveraged a carefully curated manual segmentation dataset of pediatric brain tumors and trained nnU-Net and DeepMedic models. nnU-Net showed high Dice score, sensitivity, and 95% Hausdorff distance metrics compared to DeepMedic. The distribution scale of Dice scores was also tighter with less variability across subjects for whole tumor and all tumor subregions, resulting in statistically significant higher Dice scores for nnU-Net. Additionally, for a more practical evaluation, volume percentages (proportions) of tumor subregions with respect to the whole tumor was also evaluated using core relational analysis and Bland-Altman plots, which again showed higher correlation and agreement of nnU-Net with the manual ground truth.

nnU-Net had excellent Dice scores for WT and ET, although the Dice scores for NET, CC, and ED were lower. Dice scores on a combination of all the non-enhancing areas (NET+CC+ED) was higher than these individual subregions. Our result of mean Dice scores of 0.9 for WT, 0.77 for ET, and 0.82 for the combined non-enhancing components, respectively, are comparable to the original nnU-Net results of 0.88, 0.85, 0.82, respectively for adult brain tumors that won the BraTS 2020 challenge [26]. Our results of Dice score values for both models performed worse in delineation of ED and this was particularly problematic for DeepMedic. Even though a smaller proportion of pediatric brain tumors are infiltrative compared to adult brain tumors, nevertheless the ever-present challenge of separating ED from NET persists to some degree. Overall, both models performed slightly worse on the external test cohort than on the internal test set. This was expected to some degree as the external test set was from multiple institutions with a wider variability of MR imaging acquisition protocols.

This study demonstrates the feasibility of achieving accurate pediatric brain tumor subregion segmentation results based on multiparametric, multi-scanner, multi-histology, and multi-institutional clinical standard of care MRI scans. In some cases, manual evaluation and revision may still be required in some cases to refine the segmentations, but nevertheless, the time and effort burden will be substantially less utilizing an accurate automated segmentation software.

Our study has several advantages. The MRI examinations were done as clinical standard of care scans without a predetermined universal research imaging protocol. The imaging was performed on various vendors and models of MRI scanners as well as on different field strength magnets. These features help with increasing the generalizability of the segmentation model results. The ground truth manual segmentations were also performed in a rigorous fashion through multiple iterations to ensure high-quality ground truth labeling. Segmentation evaluation studies would almost always benefit from training with even larger datasets, and this is true for our study as well. This may be particularly helpful in improving segmentation performance for external multi-institutional data where there is wider variability and scan protocols that can contribute to suboptimal performance for some tumor subregions.

The effectiveness of our approach is evidenced by the high Dice scores obtained for segmenting both the whole tumor and tumor core components using the multi-institutional BraTS-PEDs 2023 dataset. However, the lower performance in segmenting the enhancing tumor region can be primarily attributed to the very low prevalence of this subregion in subjects with DMG/DIPG tumors. Additionally, our original training cohort included a limited number of high-grade gliomas (including DMG/DIPG histology), which may have contributed to this reduced performance in the enhancing tumor region. To address this, incorporating the additional BraTS-PEDs 2023 dataset into the training data for our future autosegmentation models is a logical step. This would likely enhance the model's generalizability, particularly for the tumors with rare histology in multi-institutional cohorts.

It is worthwhile noting that no segmentation performance metric is optimal. This is exemplified by the limitations of the Dice score, as demonstrated in this study. Specifically, if a model fails to segment a certain label, the Dice score for that label will be equal to 0. This zero score significantly influences the aggregate Dice score for that label across all samples. Furthermore, if the dataset contains a limited number of samples with that particular label, the aggregate score may not accurately reflect the model's overall performance trend. As a result, we use multiple performance metrics and included tumor subregion proportion comparisons as well for a more practical and comprehensive evaluation of the segmentation results.

**CONCLUSIONS**

In summary, we present the results from automated deep learning based pediatric brain tumor subregion segmentation models from two different segmentation models, nnU-Net and DeepMedic. nnU-Net achieved excellent results and whole tumor and enhancing tumor segmentation and descent results for the nonenhancing components including nonenhancing tumor, cystic component, and peritumoral edema.

# TABLES

*Table 1 – Characteristics of patients, tumor histology, and MRI scanners in internal and external patient cohorts.*

|  | Internal Cohort | External Cohorts |
|---|---|---|
| Total Patients | 293 | 46 |
| Median age at imaging, years (range) | 7.84 (0.24 to 21.71) | 9.42 (0.55 to 20.73) |
| Sex | 157 (53.5%) Male<br>134 (45.7%) Female<br>2 (0.6%) n/a | 22 (47.8%) Male<br>24 (52.1%) Female |
| Histology |  |  |
|     Low-Grade Glioma / astrocytoma | 152 | 21 |
|     High-Grade Glioma / astrocytoma | 23 | 3 |
|     Ependymoma | 7 | 0 |
|     Medulloblastoma | 85 | 17 |
|     Brainstem glioma | 15 | 2 |
|     Germinoma | 2 | 0 |
|     Other | 9 | 3 |

*Table 2 – Results comparing the performance metrics of nnU-Net versus DeepMedic architectures for whole tumor and tumor component segmentations compared to the manual ground truth in terms of Dice score metric.*

|  | Internal Test Subjects | | External Test Subjects | |
|---|---|---|---|---|
|  | Dice score: mean±sd (median) | | Dice score: mean±sd (median) | |
| Region | nnU-Net | DeepMedic | nnU-Net | DeepMedic |
| Whole Tumor | 0.9 ± 0.07 (0.94) | 0.82 ± 0.16 (0.88) | 0.88 ± 0.07 (0.9) | 0.78 ± 0.18 (0.86) |
| Enhancing Core | 0.77 ± 0.29 (0.86) | 0.66 ± 0.32 (0.75) | 0.75 ± 0.26 (0.85) | 0.65 ± 0.32 (0.8) |
| Non-Enhancing Tumor | 0.66 ± 0.32 (0.80) | 0.48 ± 0.27 (0.49) | 0.53 ± 0.32 (0.64) | 0.4 ± 0.27 (0.4) |
| Cystic Component | 0.71 ± 0.33 (0.83) | 0.48 ± 0.36 (0.55) | 0.55 ± 0.33 (0.67) | 0.37 ± 0.33 (0.35) |
| Edema | 0.71 ± 0.40 (1) | 0.19 ± 0.33 (0) | 0.4 ± 0.43 (0.19) | 0.21 ± 0.32 (0) |
| NET+CC+ED | 0.82 ± 0.14 (0.86) | 0.67 ± 0.2 (0.73) | 0.74 ± 0.2 (0.8) | 0.59 ± 0.23 (0.65) |

*Table 3 - Results on the performance of our trained autosegmentation model using nnU-Net architecture on multi-institutional BraTS-PEDs 2023 data (n = 92). The metrics are reported for whole tumor, enhancing tumor, and tumor core (ET + CC + NET) in agreement with BraTS-PEDs evaluation method.*

| Region / Metric | Dice score: mean ± sd (median) | Sensitivity: mean ± sd (median) | 95% Hausdorff Distance: mean ± sd (median) |
|---|---|---|---|
| Whole Tumor (WT) | 0.87 ± 0.13 (0.91) | 0.82 ± 0.16 (0.88) | 6.47 ± 15.99 (3) |
| Enhancing Core (ET) | 0.48 ± 0.38 (0.58) | 0.55 ± 0.35 (0.71) | 11.26 ± 20.06 (3.40) |
| Tumor Core (TC) | 0.83 ± 0.18 (0.89) | 0.82 ± 0.17 (0.87) | 7.65 ± 16.27 (3.67) |

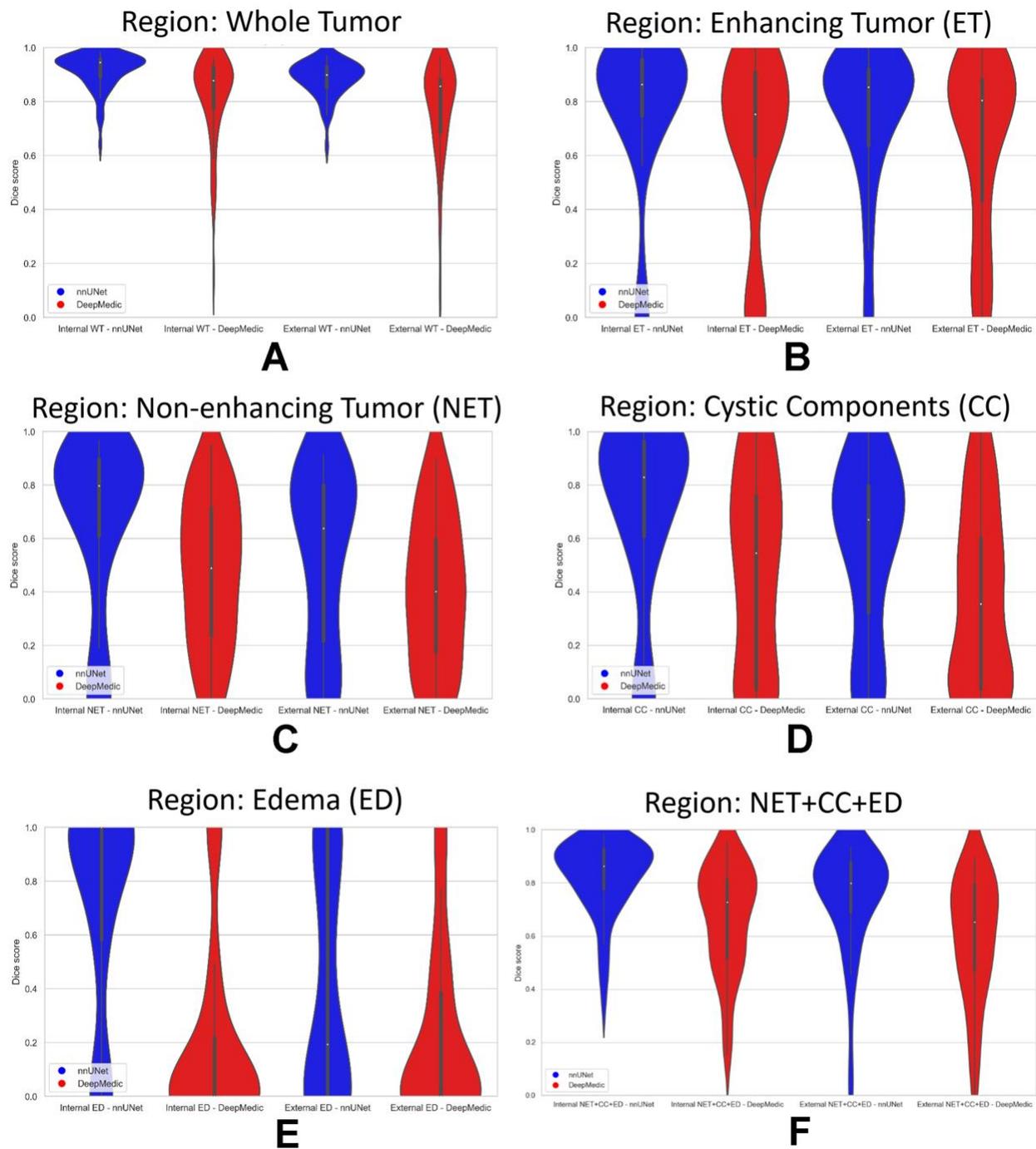

*Figure 1 – Violin plots demonstrating the distribution of Dice scores for nnU-Net and DeepMedic segmentation compared to ground truth for both internal and external test sets. (A) Whole tumor (WT). (B) Enhancing tumor (ET). (C) Nonenhancing tumor (NET). (D) Cystic component (CC). (E) Edema component (ED). (F) All nonenhancing regions (NET + CC + ED).*

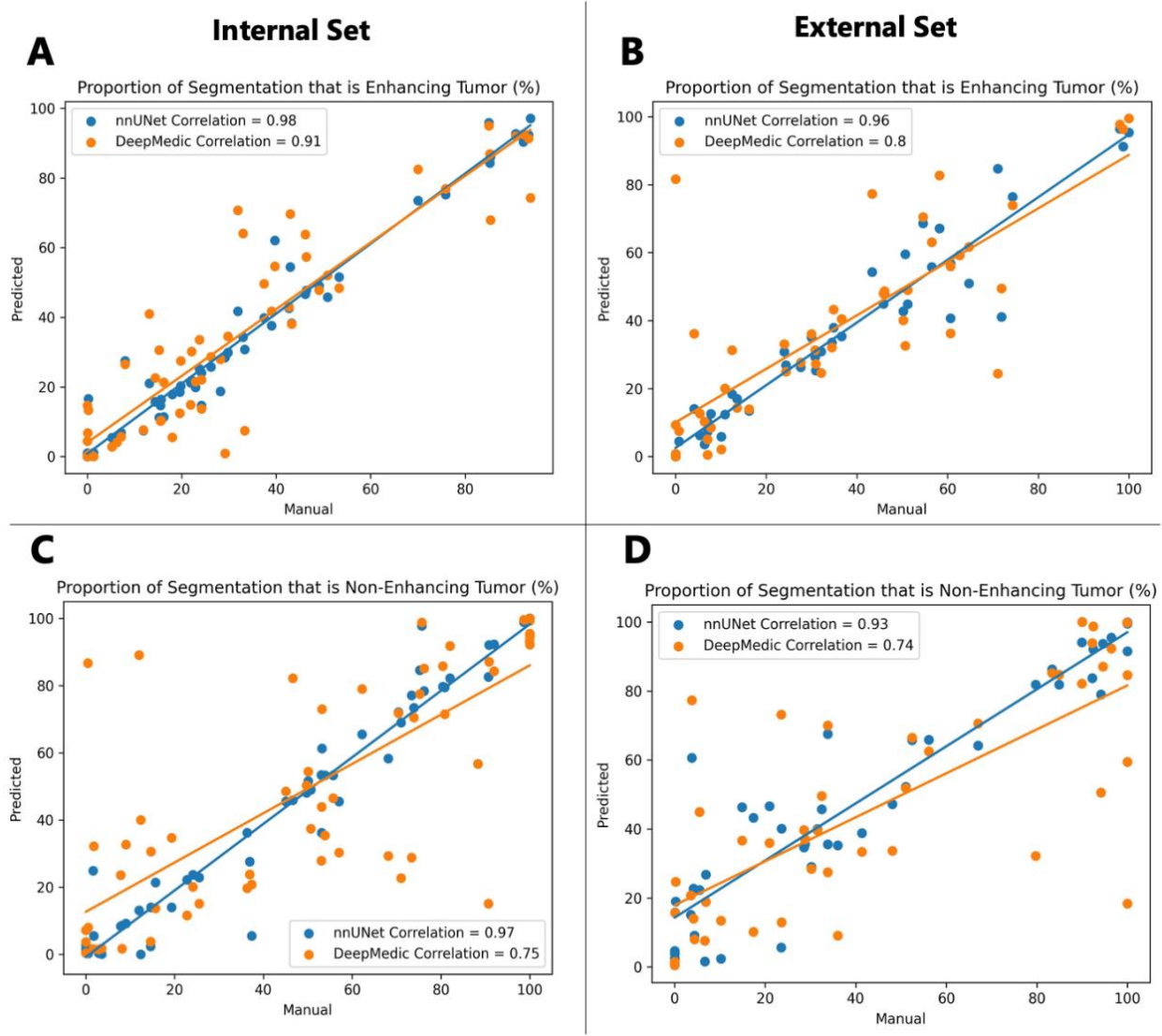

*Figure 2 – Scatterplots of the correlations between ground truth and automated tumor subregion volumes percentages from nnU-Net and DeepMedic for internal and external test sets. (A,C) proportion of tumor that is labeled enhancing; (B,D) proportion of tumor that is labeled non-enhancing tumor.*

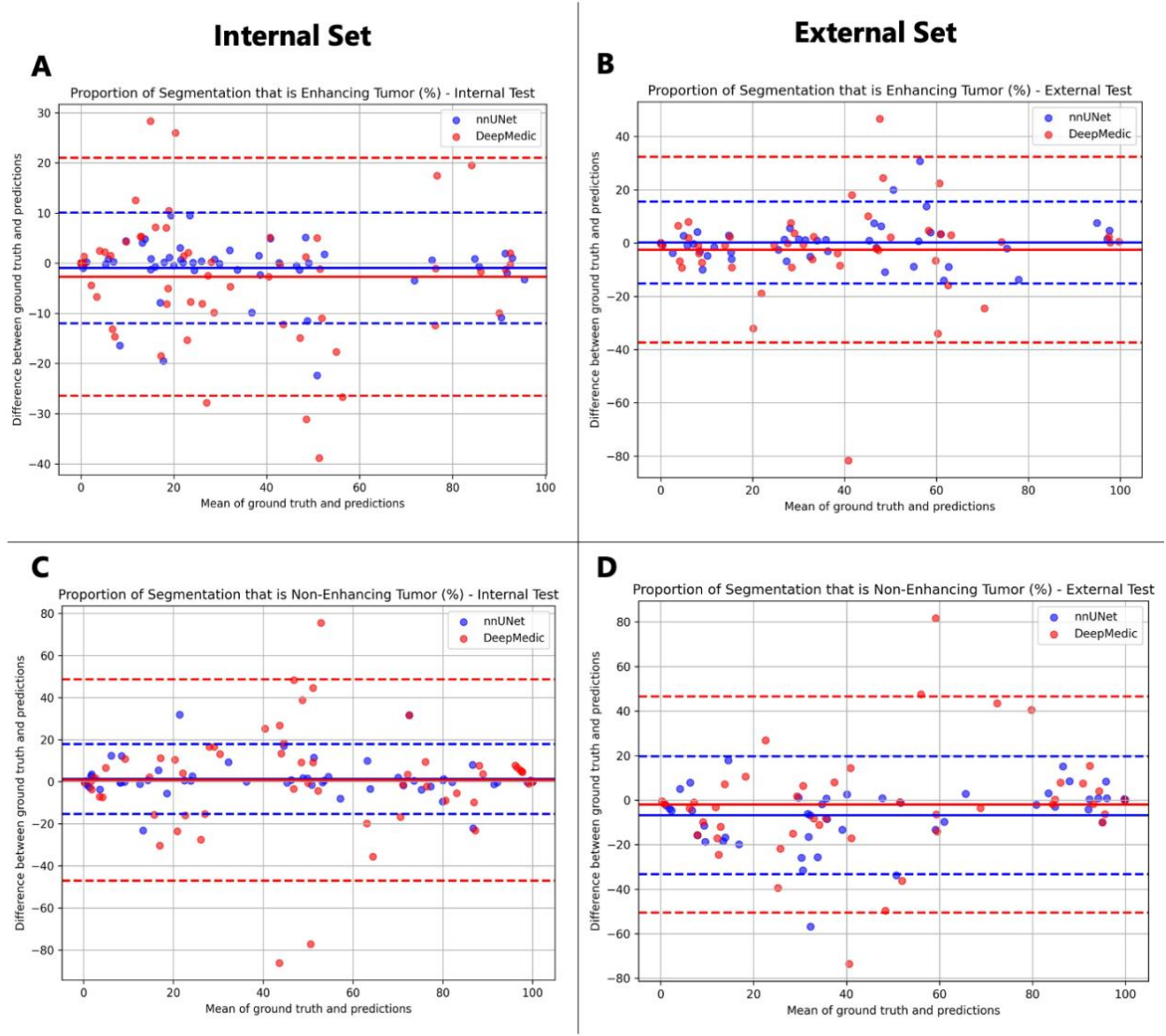

*Figure 3 – Bland-Altman analysis plots demonstrating the agreement between ground truth and automated tumor subregion volumes percentages from nnU-Net and DeepMedic for internal and external test sets. (A,C) proportion of tumor that is labeled enhancing; (B,D) proportion of tumor that is labeled non-enhancing tumor.*

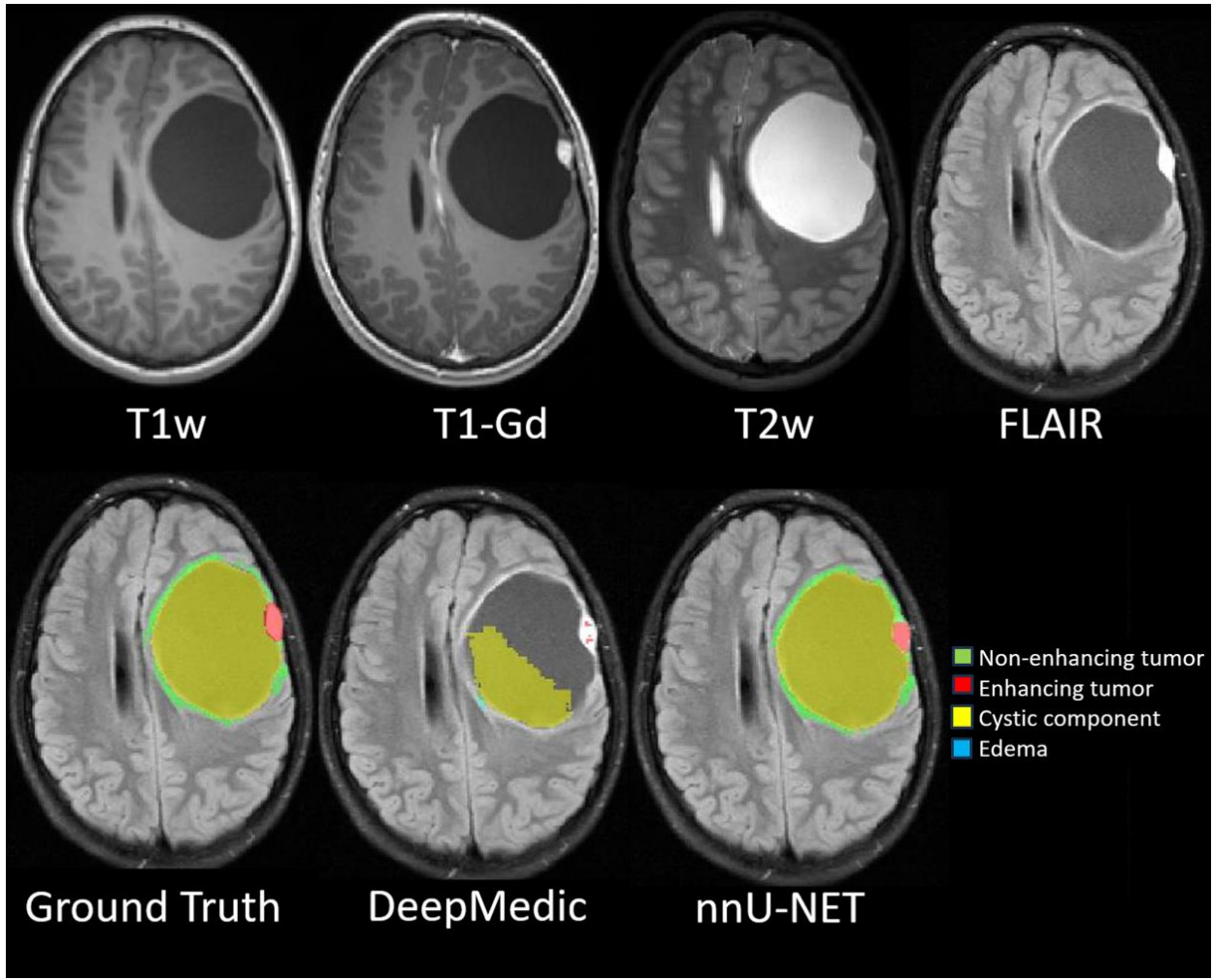

*Figure 4 – Example comparison images of ground truth and predicted segmentation from nnU-Net and DeepMedic in a predominantly cystic supratentorial mass with a solid enhancing nodule. While nnU-Net has near perfect agreement with the ground truth, DeepMedic failed to delineate most of the large cyst, enhancing nodule, and nonenhancing rim.*

**Supplementary Material**

*Image Preprocessing*

The MRI scans for each patient were first re-oriented to left-posterior-superior (LPS) coordinate system. Subsequently, T1WI, T2WI, and FLAIR sequence images were co-registered with their corresponding T1WI-Gd sequence and resampled to an isotropic resolution of 1 mm$^3$ based on the anatomical SRI24 atlas [1] using a greedy algorithm in the Cancer Imaging Phenomics Toolkit open-source software v.1.8.1 (CaPTk, https://www.cbica.upenn.edu/captk) [2]. Initial preliminary brain tumor subregion segmentation into enhancing tumor (ET), non-enhancing tumor (NET), cystic component (CC), and peritumoral edema (ED) was performed using an automated pediatric brain tumor segmentation tool [3]. Subsequently, detailed manual revisions on this preliminary segmentation were made by trained students and non-radiologist physician researchers in ITK-SNAP [4]. Finally, revisions were reviewed and finalized by one of three neuroradiologists (J.W. with 6, A.N. with 10, and A.V. with 16 years of neuroradiology experience, respectively). Some tumors were manually revised through multiple iterations. All manual annotators, including the neuroradiologists, had undergone previous training and harmonization consensus sessions. In some challenging cases, such as in determination of presence of subtle contrast enhancement or edema delineation, the images were reviewed by all three neuroradiologists to reach consensus. The finalized manual segmentations were used as ground truth for model training and evaluation.

*Supplementary Table 1 – Characteristics of MRI scanners in internal and external patient cohorts.*

|  | Internal Cohort | External Cohorts |
|---|---|---|
| Total Patients | 293 | 46 |
| Scanner Magnetic Field Strength (Tesla) | | |
|    0.7 | 0 | 1 |
|    1.5 | 75 | 35 |
|    3 | 218 | 10 |
| Scanner Manufacturer | | |
|    Siemens | 271 | 26 |
|    GE | 17 | 15 |
|    Phillips | 5 | 4 |
|    Toshiba | 0 | 1 |

*Supplemental Table 2 – Results comparing the performance metrics of nnU-Net versus DeepMedic architectures for whole tumor and tumor component segmentations compared to the manual ground truth in terms of Sensitivity metric.*

| Region | Internal Test Subjects | | External Test Subjects | |
|---|---|---|---|---|
| | Sensitivity: mean±sd (median) | | Sensitivity: mean±sd (median) | |
| | nnU-Net | DeepMedic | nnU-Net | DeepMedic |
| Whole Tumor | 0.9 ± 0.09 (0.93) | 0.80 ± 0.19 (0.86) | 0.86 ± 0.1 (0.86) | 0.8 ± 0.18 (0.85) |
| Enhancing Core | 0.78 ± 0.24 (0.84) | 0.67 ± 0.30 (0.74) | 0.78 ± 0.21 (0.86) | 0.74 ± 0.26 (0.83) |
| Non-Enhancing Tumor | 0.69 ± 0.28 (0.79) | 0.52 ± 0.26 (0.46) | 0.61 ± 0.28 (0.71) | 0.49 ± 0.28 (0.52) |
| Cystic Component | 0.71 ± 0.26 (0.80) | 0.49 ± 0.28 (0.49) | 0.51 ± 0.28 (0.61) | 0.36 ± 0.29 (0.34) |
| Edema | 0.46 ± 0.37 (0.62) | 0.15 ± 0.18 (0.08) | 0.19 ± 0.26 (0.01) | 0.21 ± 0.23 (0.13) |
| NET+CC+ED | 0.81 ± 0.17 (0.87) | 0.65 ± 0.22 (0.69) | 0.74 ± 0.16 (0.78) | 0.63 ± 0.22 (0.67) |

*Supplemental Table 3 – Results comparing the performance metrics of nnU-Net versus DeepMedic architectures for whole tumor and tumor component segmentations compared to the manual ground truth in terms of 95% Hausdorff Distance metric.*

|  | Internal Test Subjects | | External Test Subjects | |
|---|---|---|---|---|
|  | 95% Hausdorff Distance: mean±sd (median) | | 95% Hausdorff Distance: mean±sd (median) | |
| Region | nnU-Net | DeepMedic | nnU-Net | DeepMedic |
| Whole Tumor | 3.79 ± 9.25 (2) | 15.42 ± 18.99 (5.19) | 3.79 ± 9.25 (2) | 15.42 ± 18.99 (5.19) |
| Enhancing Core | 4.06 ± 6.15 (2) | 6.89 ± 7.86 (4.06) | 4.06 ± 6.15 (2) | 6.89 ± 7.86 (4.06) |
| Non-Enhancing Tumor | 8.01 ± 14.02 (2.28) | 17.02 ± 17.31 (7.21) | 8.01 ± 14.02 (2.28) | 17.02 ± 17.31 (7.21) |
| Cystic Component | 5.20 ± 4.84 (3.39) | 14.67 ± 17.44 (8.60) | 5.20 ± 4.84 (3.39) | 14.67 ± 17.44 (8.60) |
| Edema | 14.60 ± 18.64 (6.40) | 25.81 ± 15.02 (20.22) | 14.60 ± 18.64 (6.40) | 25.81 ± 15.02 (20.22) |
| NET+CC+ED | 4.37 ± 9.31 (2.24) | 13.79 ± 18.42 (5.39) | 4.37 ± 9.31 (2.24) | 13.79 ± 18.42 (5.39) |

*Supplemental Table 4 – Pearson's correlation coefficients between predicted and ground truth manual segmentation volumes of Enhancing Tumor, Non-Enhancing Tumor, and Cystic Component subregions, and Spearman's rank correlation coefficients for the Edema region. All P values were less than 0.001.*

| Portion of tumor | Internal Test Subjects | | External Test Subjects | |
|---|---|---|---|---|
| | nnU-Net vs Ground Truth | DeepMedic vs Ground Truth | nnU-Net vs Ground Truth | DeepMedic vs Ground Truth |
| Enhancing Tumor | 0.98 | 0.91 | 0.96 | 0.8 |
| Non-Enhancing Tumor | 0.97 | 0.75 | 0.93 | 0.74 |
| Cystic Component | 0.98 | 0.8 | 0.93 | 0.7 |
| Edema | 0.95 | 0.48 | 0.88 | 0.33 |

*Supplementary Table 5 - Results of Mann-Whitney U test (p-values) comparing tumor subregion as a proportion of whole tumor in nnU-Net and DeepMedic segmentations compared to ground truth segmentation.*

|  | Internal Test Subjects | | External Test Subjects | |
|---|---|---|---|---|
| Portion of Tumor | nnU-Net vs Ground Truth | DeepMedic vs Ground Truth | nnU-Net vs Ground Truth | DeepMedic vs Ground Truth |
| Enhancing Tumor | 0.83 | 0.64 | 0.99 | 0.6 |
| Non-Enhancing Tumor | 0.83 | 0.91 | 0.28 | 0.62 |
| Cystic Component | 0.7 | 0.44 | 0.64 | 0.62 |
| Edema | 0.95 | 0.003* | 0.46 | 0.1 |

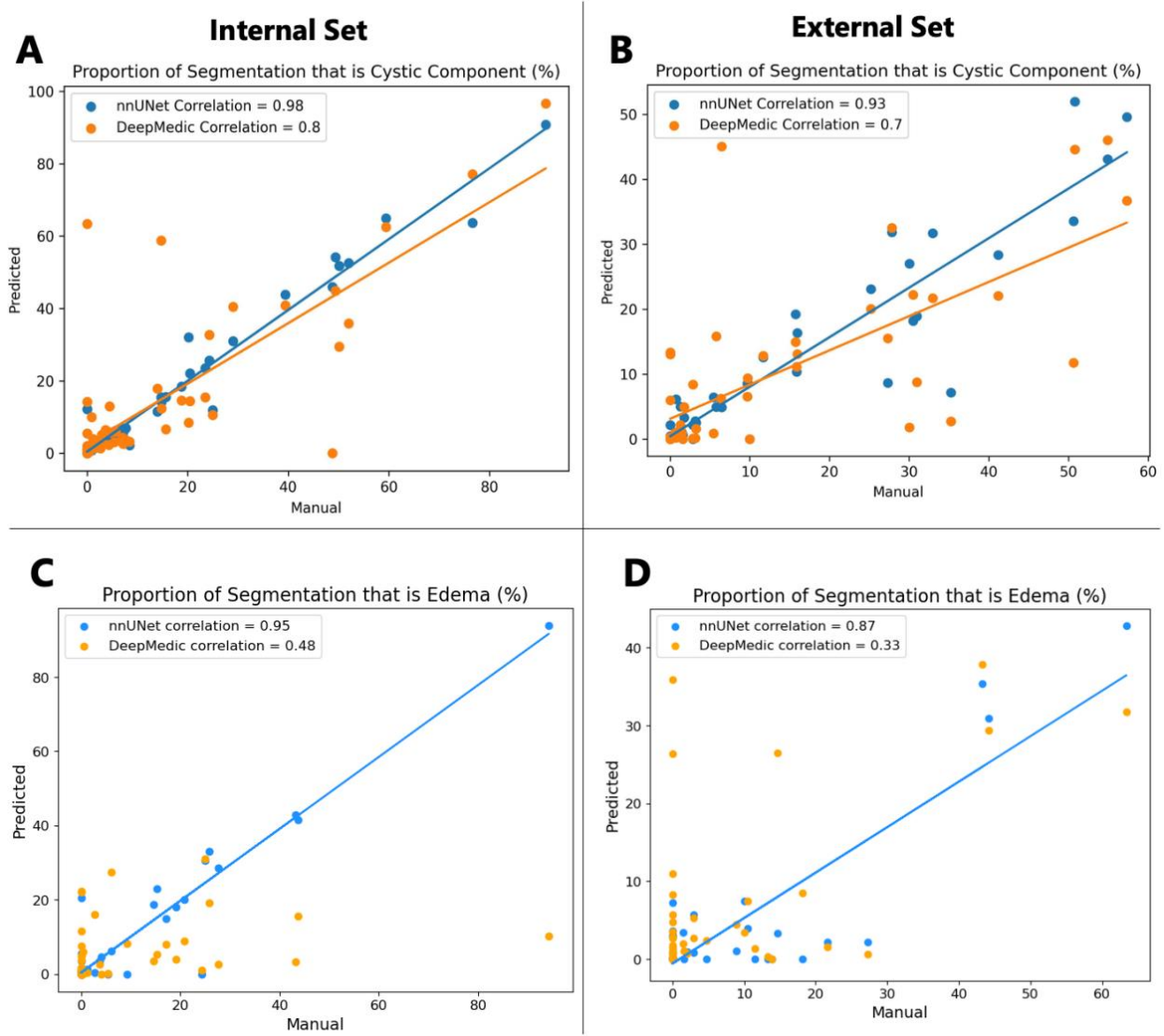

*Supplementary Figure 1 – Scatterplots of the correlations between ground truth and automated tumor subregion volumes percentages from nnU-Net and DeepMedic for internal and external test sets. (A,C) proportion of tumor that is labeled cystic component; (B,D) proportion of tumor that is labeled edema. The scatter plots for the predictions made by DeepMedic method within the edema region do not include fitted lines, as the correlations were calculated using the non-parametric Spearman's rank correlation.*

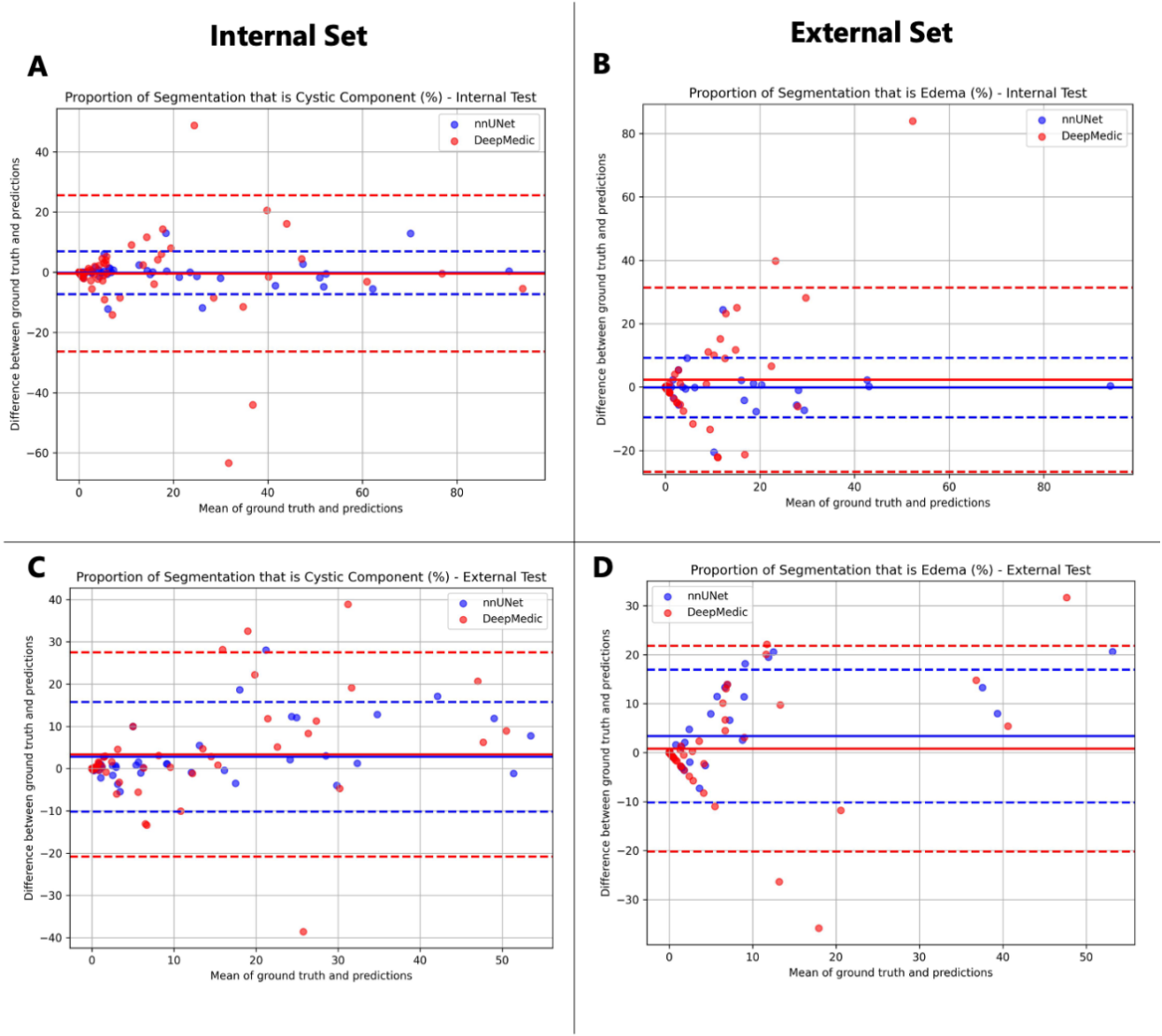

*Supplementary Figure 2 – Bland-Altman analysis plots demonstrating the agreement between ground truth and automated tumor subregion volumes percentages from nnU-Net and DeepMedic for internal and external test sets. (A,C) proportion of tumor that is labeled cystic component; (B,D) proportion of tumor that is labeled edema.*

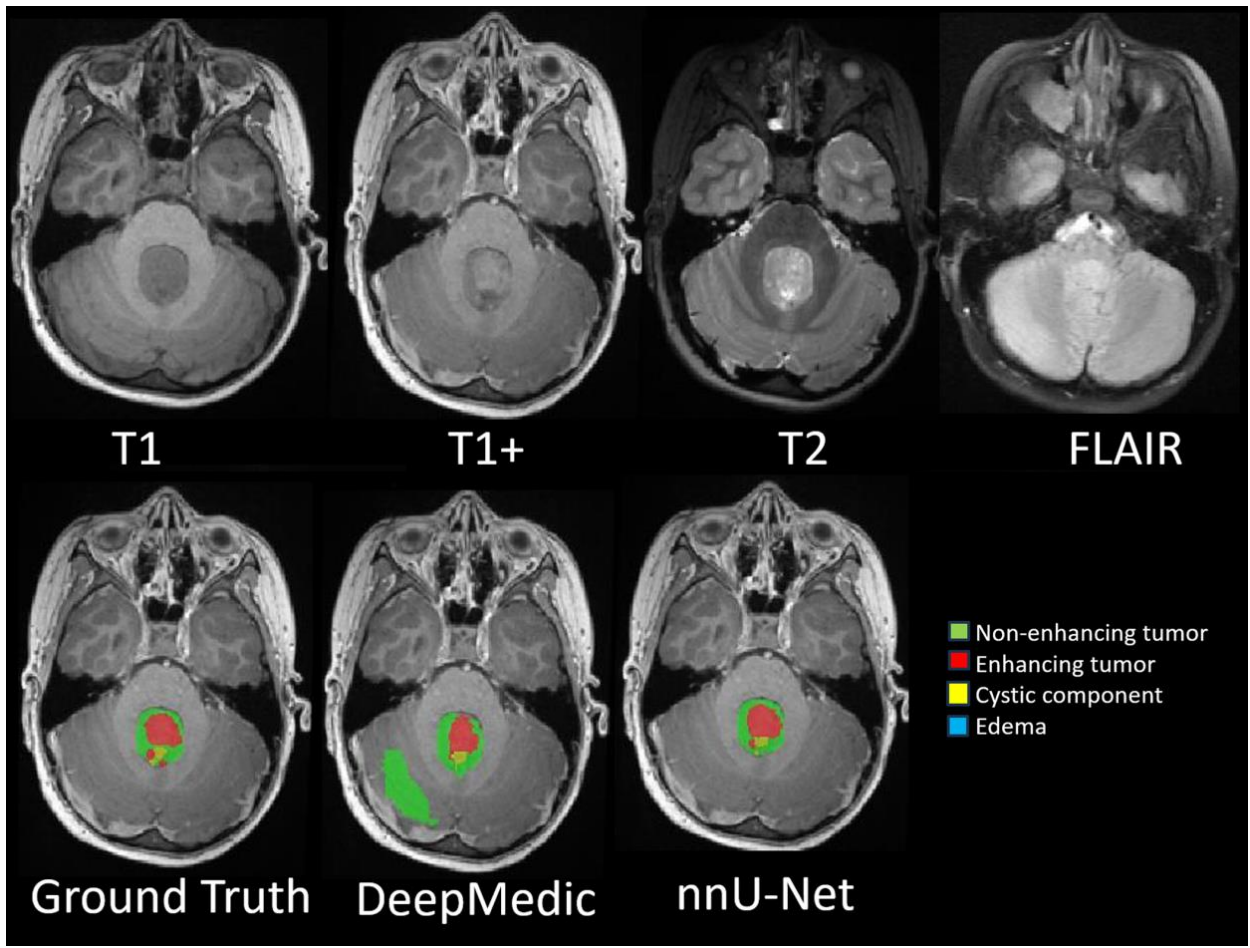

*Supplementary Figure 3 – Example comparison images of ground truth and predicted segmentation from nnU-Net and DeepMedic in a posterior fossa medulloblastoma. While both DeepMedic and nnU-Net performing well in tumor region segmentation compared to the ground truth albeit nnU-Net more closely mimicking ground truth segmentation, DeepMedic demonstrates an area of false-positive nonenhancing tumor prediction in the right cerebellar hemisphere.*

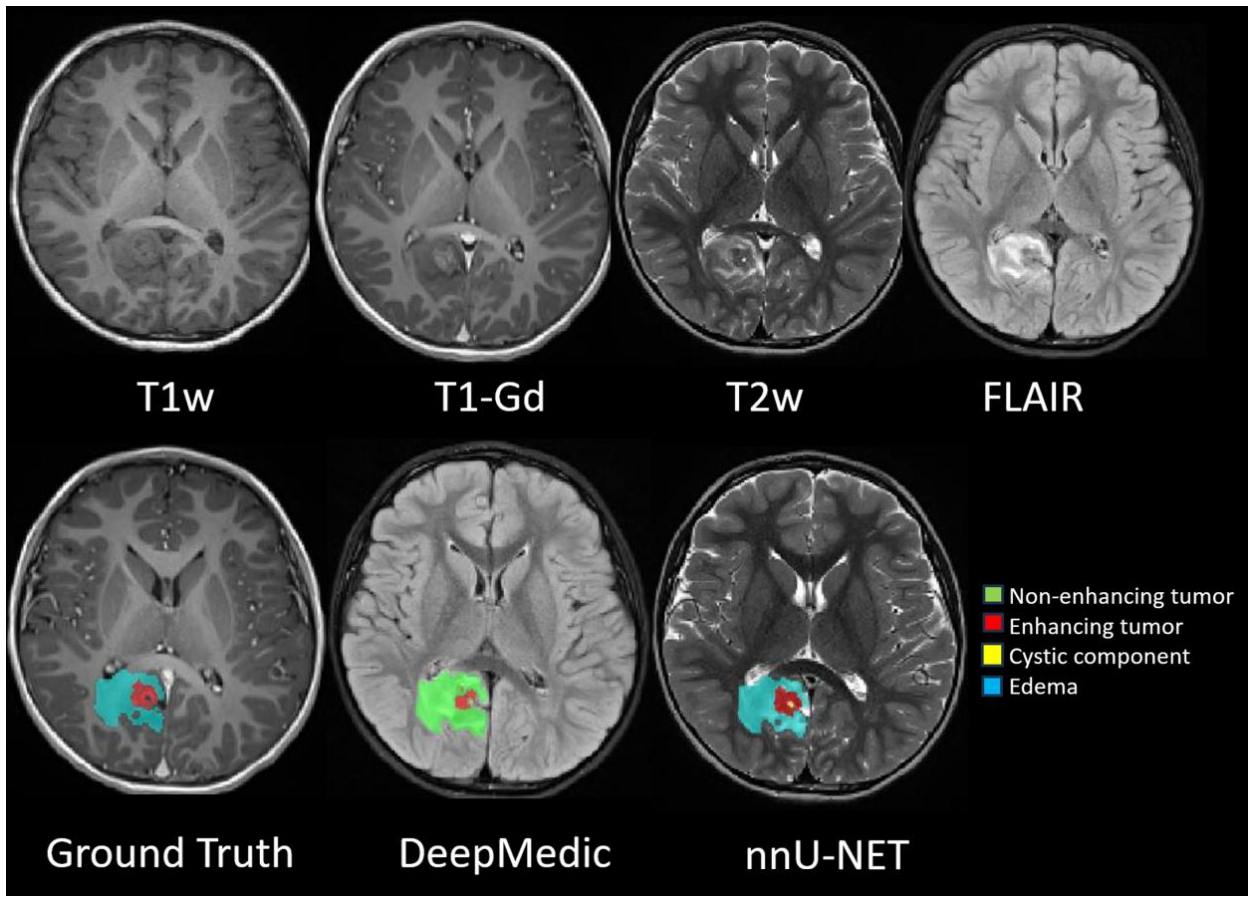

*Supplementary Figure 4 – Example comparison images of ground truth and predicted segmentation from nnU-Net and DeepMedic in a subject with an enhancing right occipital parietal glioma. Although the whole tumor segmentation is well demarcated by both models, DeepMedic is erroneously labeling edema as nonenhancing tumor.*

**Supplementary Material References**